\documentclass[12pt, a4paper]{article}

\usepackage{amsmath}
\usepackage{amsfonts}
\usepackage{amssymb}
\usepackage{graphicx, rotating}
\usepackage{epstopdf}
\usepackage{epsfig}
\usepackage{latexsym}
\usepackage{graphicx}
\usepackage{color}
\usepackage{amsmath,bm,amssymb}
\usepackage{cite}
\usepackage{slashed}
\usepackage{hyperref}
\hypersetup{colorlinks, citecolor=bluscuro, linkcolor=black, urlcolor=bluscuro}
\definecolor{rossos}{cmyk}{0,1,1,0.55}
\definecolor{bluscuro}{rgb}{0.15, 0.2, .85}
\definecolor{bluchiaro}{cmyk}{1,.3,0.,0.1}
\definecolor{brown}{rgb}{0.6, 0.14, 0.14}

\def\simlt{\stackrel{<}{{}_\sim}}

\newcommand{\be}{\begin{equation}}
\newcommand{\ee}{\end{equation}}
\newcommand{\bea}{\begin{eqnarray}}
\newcommand{\eea}{\end{eqnarray}}

\begin{document}

%FRONTPAGE2%%%%%%
\begin{titlepage}
\begin{flushright}
%UAB-FT--\\
%CERN-PH-TH/2013-029
\end{flushright}
\vspace{.3in}

\vspace{1cm}
\begin{center}
{\Large\bf\color{black} Higgs Inflation as a Mirage}\\

\bigskip\color{black}
\vspace{1cm}{
{\large J.~L.~F. Barb\'on$^a$, J.~A. Casas$^a$,  J.~Elias-Mir\'o$^{b,c}$, J.R.~Espinosa$^{b,d}$}
\vspace{0.3cm}
} \\[7mm]
{\it {$^a$\, IFT-UAM/CSIC, Madrid}}\\
{\it {$^b$\, IFAE, Universitat Aut{\`o}noma de Barcelona,
   08193~Bellaterra,~Barcelona}}\\
{\it {$^c$\, Dept.~de~F\'isica, Universitat Aut{\`o}noma de Barcelona, 08193~Bellaterra,~Barcelona}}\\
{\it $^d$ {ICREA, Instituci\'o Catalana de Recerca i Estudis Avan\c{c}ats, Barcelona, Spain}}
\end{center}
\bigskip

\vspace{.4cm}

\begin{abstract}
We discuss a simple unitarization of Higgs inflation that is genuinely weakly coupled up to Planckian energies. A large
non-minimal coupling between the Higgs and the Ricci curvature is induced dynamically at intermediate energies, as a simple ratio of mass scales. Despite not being dominated by the Higgs field, inflationary dynamics simulates  the `Higgs inflation'
one would get by blind extrapolation of the low-energy effective Lagrangian, at least qualitatively. Hence, Higgs inflation arises as an approximate  `mirage'  picture of the true dynamics. We further speculate on the generality of this phenomenon and show that, if Higgs-inflation arises as an effective description, the details of the UV completion are necessary to extract robust quantitative predictions.

\end{abstract}
\bigskip

\end{titlepage}

%%%%%%%%%%%%%%%%%%%%%%%%%%%%%%%%%%%%%%%%
%%%%%%%%%%%%%%%%%%%%%%%%%%%%%%%%%%%%%%%%
%%%%%%%%%%%%%%%%%%%%%%%%%%%%%%%%%%%%%%%%
%%%%%%%%%%%%%%%%%%%%%%%%%%%%%%%%%%%%%%%%

\section{Introduction}

\noindent

The construction of an inflationary model often begins by postulating a scalar field with the right properties to drive an approximately de Sitter phase of the universe, followed by a  graceful exit into a hot big bang. Just as it happens in other instances of model building, there is a balance to strike between the desired economy of  degrees of freedom and the need to account for possibly intricate dynamics. 

 Taking the minimalistic attitude to the extreme, we could entertain the
possibility that one and the same Standard Model (SM)  Higgs field may control {\it both}  the vacuum properties {\it and} the inflationary dynamics in different corners of its configuration space.  As shown  in \cite{shap}, this feat can be achieved with stunning simplicity. It suffices to add one extra parameter to the SM, associated to a non-minimal Higgs-graviton interaction, consisting on a Higgs mass term proportional to the background scalar curvature:  
\begin{equation}
\label{nonmin}
{\cal L}_\xi = -\xi\,|H|^2 \,R
\;.
\end{equation}
The trick works 
provided the dimensionless coupling can be tuned to be parametrically  large, $\xi \gg 1$.  This so-called `Higgs-inflation' (HI) scenario operates at large field values, $|H| >  M_p / \sqrt{\xi}$, a  regime in which gravity is weakened in proportion to  a field-dependent Newton's constant given by $(M_p^2 + 2\xi \langle |H|^2 \rangle)^{-1}$. In this situation, it is common practice to work in the equivalent   `Einstein frame', obtained by an appropriate Weyl rescaling of the metric. In the Einstein frame one deals with a fixed Newton constant and non-minimal couplings inside covariant derivatives.  The dependence on $\xi$ is transferred to a tower of effective operators which change the dynamics of the Higgs modulus field for $|H| >  M_p /\xi$.  
The most visible of these changes is a rescaling of the Higgs potential $U(H)=\lambda |H|^4$ into 
 \be\label{ppot}
V(H) = {\lambda |H|^4 \over \left(1+ 2\xi {|H|^2 \over M_p^2}\right)^2} 
\;,
\ee
 which features an asymptotic plateau extending beyond $|H| \sim  M_p /\sqrt{\xi}$,  with energy density $V_\infty = \lambda M_p^4 /4\xi^2$. This is the key to inflation in this model. The inflationary dynamics implied by (\ref{ppot}) is robust, since the Weyl transformation has the additional effect of  decoupling the Higgs modulus from the rest of the SM degrees of freedom, precisely in the  `plateau region' of field space, $|H|\gg M_p /\sqrt{\xi}$. 
 
  Detailed analysis reveals that $\xi$ must be chosen in the ballpark of $10^4$ to fit the correct amplitude of cosmological perturbations, although quantum corrections to the effective potential have a significant impact on this fit value. In particular, the top Yukawa coupling brings down the Higgs self-coupling at high energies and, more generally, the running of the various marginal couplings introduces a logarithmic sensitivity of inflationary physics on low-energy parameters \cite{shap,HImore}. These effects result in significantly smaller fit values for $\xi$, but they also introduce a tension with the measured values of the Higgs and top quark masses. 
 
The scenario of Higgs inflation that we have described is in serious disagreement with the `standard rules' of effective field theory, since it was  defined by a bold extrapolation of an effective Lagrangian beyond its naive domain of applicability. When analyzed near the SM vacuum, the operator (\ref{nonmin}) has dimension five by power counting, with an effective cutoff scale 
$$
\Lambda \equiv {M_p \over \xi}\;,
$$ 
well below the edge of the plateau \cite{burgess,cutoff}.\footnote{The operator in Eq.~(\ref{nonmin}) is actually redundant if standing alone, since it can be removed by the Weyl rescaling combined with a further field redefinition of the scalar field. However, any  ${\cal O}(1)$ interactions of $|H|$, either with itself or any other degrees of freedom, activate $\Lambda$ as a physical  dynamical scale
\cite{cutoff}.} 
 
 The very existence of an asymptotic plateau in the extrapolation depends on the precise correlation between the original Higgs potential $U(|H|)$ and the non-minimal coupling, which must be proportional to $\sqrt{U(|H|)}$ at large values of $|H|$. One way of making this assumption explicit is to declare the existence of an asymptotic shift symmetry acting on the Higgs modulus when the theory is written in the  Einstein frame. This asymptotic shift symmetry should be interpreted as a property of some  hypothetical  UV completion of the model. 

In principle, the existence of an asymptotic shift symmetry, with its associated Einstein-frame plateau, is independent of the  $\xi \gg 1$ condition. Still, a parametrically large value of $\xi$ does play an important role in the structure of the model. First, it is required to fit the amplitude of inflationary perturbations. Second, it is required for consistency of the semiclassical approximation during slow-roll, ensuring that curvatures remain sub-Planckian during inflation. To see this, notice that the Hubble constant during inflation is given by $H_I^2 \sim V_\infty /M_p^2$, where $V_\infty$ is the vacuum energy density along the plateau. Therefore we have
$$
\left({H_I \over M_p}\right)^2 \sim {\lambda \over \xi^2}
\;,
$$
so that the consistency of the  effective field theory of inflation requires $\lambda/\xi^2 \ll 1$. 
 Although the SM running of $\lambda$ towards small values  helps in enforcing this inequality, this argument shows that a parametrically large value of $\xi$ is  often a useful ingredient in the construction of these models. 

On the other hand, a large value of $\xi$ introduces a 
hierarchy between the naive cutoff scale $\Lambda$ and the edge of the inflationary plateau, lying  at field values of ${\cal O}(\sqrt{\xi} \Lambda)$. This  intermediate region in the  Higgs configuration space, 
\be\label{intreg}
\Lambda \ll |H| \ll \sqrt{\xi} \,\Lambda\;,
\ee
 is characterised by large corrections to the kinetic metric of  $|H|$, to the point of compromising the unitarization of longitudinal $W$-boson interactions --the very {\it raison d'\^etre} of the SM Higgs-- in any background with $|H| > \Lambda$. Hence, the introduction of extra degrees of freedom at the scale $\Lambda$ is unavoidable. This new physics beyond the SM has the potential to affect the Higgs-gravity dynamics, threatening the consistency of the Higgs inflation scenario.    In other words, the low-energy physics at large values of $|H|$, as inferred from the UV-completed model, is bound to differ in every detail from the naive extrapolation operating in the so-called `Higgs inflation' scenario.  

The purpose of this paper is to introduce a simple model which allows us to  separate the  actual inflationary mechanism, which operates at very large fields, from the specific problem posed by the existence of the intermediate hierarchy in Eq.~(\ref{intreg}). The model amounts to a simple extension of the SM by an extra scalar singlet  with  carefully chosen interactions. 
 
 A crucial property of the model is the absence of any large irrelevant operators below $M_p$ in the effective UV description. The  operator in Eq.~(\ref{nonmin})  is dynamically  induced  at low energies,  with  $\xi \gg 1$,    by the interplay of {\it relevant} operators acting at intermediate scales. 
In this respect, the unnatural choice  $\xi \gg 1$ has  the same character as the familiar violation of naturalness present in the SM, in the choice of the Higgs mass parameter. Unlike the case of other UV completions such as \cite{gianlee}, there are no large irrelevant operators in the UV effective theory, so that all visible energy scales appear as weakly coupled thresholds below the Planck scale. 

Viewed as an extrapolated Higgs-inflation scenario, our model is not exactly identical to \cite{shap}: it does produce  slightly different values for the main inflationary observables, such as the spectral index  and the tensor/scalar ratio, but it certainly falls under the same qualitative category of models. Rather than targeting realistic phenomenology, the main purpose of our exercise is to extract general lessons regarding the reliability of blind extrapolations of effective Lagrangians with large non-minimal couplings of type (\ref{nonmin}). Our analysis indicates that extrapolations are likely to behave as a `mirage' of the true inflationary dynamics, capturing the gross qualitative features but offering a `blurred' picture when it comes to the details.

The plan of the paper is as follows. In section 2 we introduce the basic structure of the UV model and its low-energy approximation. In section 3 we proceed to study the behaviour of both models when they are extrapolated to the large field regime, including a discussion of the properties of inflation in the UV theory. Finally, section 4 is devoted to the interpretation of our results in the light of effective field theory.

\section{The Model} 

We consider a simple extension of the Standard Model with one single new degree of freedom: a heavy (real) scalar $\phi$ with a (Jordan-frame) Lagrangian given by\footnote{The Jordan frame is defined by a standard minimal coupling in covariant derivatives, and a possibly field-dependent Newton's constant. Conversely, the Einstein frame is defined by a field-independent Newton's constant and generally non-minimal couplings in covariant derivatives. The two are related by a  Weyl rescaling of the metric.}  
\be
{\cal L}_{\rm Jordan}= \left[
-\frac{1}{2}M_p^2 \,R - g M_p \,\phi \,R +\frac{1}{2}(\partial_\mu \phi)^2 -U(\phi,H)\right]+{\cal L}^{SM}\ ,
\label{LJUV}
\ee
where ${\cal L}^{SM}$ is the Standard Model Lagrangian with minimal coupling to the space-time metric. The scalar $\phi$ has a linear non-minimal coupling to  the Ricci curvature scalar, with a dimensionless strength $g$.  Most of our discussion will take place under the assumption that $g ={\cal O}(1)$, but we will find it useful to consider other dynamical regimes, involving the tuning of $g$ to extreme values, large or small. The scalar potential involving $\phi$,
\be
U(\phi,H)=\frac{1}{2}m^2\,\phi^2 -\mu\, \phi\, |H|^2\ ,
\ee
introduces two  mass scales $\mu$ and $m$, while the SM Higgs potential introduces a further energy scale $\mu_h$ and reads
\be\label{smpot}
U^{SM}(H) = - \mu_h^2 \,|H|^2 + \lambda_0\, |H|^4\ .
\ee
We do not impose any naturalness constraints on these mass parameters. By this we mean that we allow large hierarchies   between all mass scales by explicit tuning of dimensionless ratios, such as $m /M_p$, $\mu_h / m$.  This violation of naturalness is to be interpreted as a generalisation of the standard tuning of $\mu_h /M_p$ in the minimal SM. 

The compatibility with low-energy SM phenomenology imposes some constraints on the mass parameters. For instance, the trilinear coupling $\mu$ 
 cannot be too large as it contributes to $\phi-h$ mixing in the electro-weak (EW) vacuum. More explicitly,
the EW vacuum  determined by the above potential sits at
\be
\langle h\rangle ^2= \frac{\mu_h^2}{\lambda} = v^2\ , \quad
\langle\phi\rangle = \frac{1}{2}\frac{\mu v^2}{m^2}\ ,
\ee
where $h$ denotes the (real) neutral Higgs component and  we have introduced
\be
\lambda \equiv \lambda_0 - \frac{1}{2}\frac{\mu^2}{m^2}. 
\label{lambdaeff}
\ee
Assuming that the coupling $\lambda_0$ is perturbative, we find an upper bound on how large $\mu$ can be, with $\mu^2\leq 2\lambda_0 m^2\simlt m^2$.
The mass matrix for the $CP$-even scalars $\{h,\phi\}$ reads
\be
{\cal M}^2 = \left[
\begin{array}{cc}
-\mu_h^2 + 3\lambda_0 h^2-\mu \phi & -\mu h\\
-\mu h & m^2
\end{array}
\right] =
\left[
\begin{array}{cc}
2\lambda v^2 + \mu^2 v^2/m^2 & -\mu v\\
-\mu v & m^2
\end{array}
\right] 
\ ,
\ee
where the second expression holds at the minimum.
In an expansion in powers of $v/m$, the two mass eigenvalues are
\be
m_h^2 \simeq 2\lambda v^2 + {\cal O}(\mu^2 v^4/m^4) \ , \quad
m_\Phi^2 \simeq m^2 + \frac{\mu^2 v^2}{m^2}+   {\cal O}(\mu^2 v^4/m^4) \ .
\ee
The first corresponds to the SM Higgs and the second to a heavy singlet. Notice in particular that the mixing angle is of order $\mu v/m^2 \ll 1$ so that the light state has SM properties.

Turning our attention to the high energy physics, we notice that the Lagrangian (\ref{LJUV}) is far from being  generic at the level of marginal and irrelevant couplings: we have engineered the complete absence of $\phi$ self-interactions,  and the non-miminal coupling $g\, \phi\, R$ is the only irrelevant operator containing the new scalar field $\phi$.  Therefore, the Lagrangian (\ref{LJUV}) defines a weakly-coupled effective field theory all the way up to the Planck scale, provided $g\leq 1$, a condition that we will regard as the   `unitarity constraint'.  In this case, all mass scales below $M_p$ are explicitly visible in the Lagrangian, with no hidden strong-coupling thresholds.

The concrete scenario of inflation does depend on the particular value of the non-minimal coupling. 
For $g\ll 1$  we have a garden-variety model of chaotic $\phi^2$-inflation.  This region of parameter space will not be our main concern in this paper, although it will serve as an instructive toy model in section 4. 

Our main interest is the case with $g$ of ${\cal O}(1)$, where the non-minimal coupling becomes strong precisely at the Planck scale. The precise correlation between the non-minimal coupling, $\phi\,R$, and the highest power of the bare potential, $\phi^2$, allows us to run a rehash of the Higgs inflation scenario, by simply extrapolating the Einstein-frame effective Lagrangian to the positive trans-Planckian domain\footnote{Regarding extrapolations, we must stay clear of the negative trans-Planckian region, since the effective Newton constant of the Jordan frame becomes negative for $\phi <-M_p /2g$. We will see below that this pathology is milder than it appears to be. Nevertheless, it does restrict the scope of (\ref{LJUV}) as an effective theory in field space. }
$\phi \gg M_p$. A crucial difference with the standard HI scenario is the absence of any large dimensionless couplings in the action, particularly in the non-minimal coupling of type (\ref{nonmin}). 

A radical violation of the unitarity constraint, in the form of a large non-minimal coupling $g\gg 1$, would just reintroduce the large-$\xi$ problem of HI into the effective UV model (\ref{LJUV}).  Even if such a consideration runs against the main philosophy of this work, it is worth mentioning that the formal limit $g\rightarrow \infty$ can be analysed from the standpoint of (\ref{LJUV}) by rescaling both the field $\phi \rightarrow \Phi/g$ and the mass parameter $m \rightarrow M/g$. In the new variables, taking  $g\rightarrow \infty$ removes the kinetic action for the  $\Phi$ field, which turns into an ordinary Lagrange multiplier. In this limit   (\ref{LJUV}) becomes equivalent to Starobinsky's model of inflation \cite{starob}, with mass scale~$M$. 

\subsection{Low energy effective action}

The simple choice of action in  (\ref{LJUV}) allows us to integrate out the field $\phi$ exactly, leading to the formal expression  
\be\label{forme}
{\cal L}_{\rm eff} = {\cal L}^{SM} -{1\over 2}M_p^2\, R + {1\over 2} \left(\mu\,|H|^2 - g M_p\,R\right) {1\over m^2 + \square} \left(\mu\,|H|^2 - g M_p\,R\right) +\dots
\ee
where the dots stand for the one-loop effective action, proportional to ${\rm Tr} \,\log\,(\square + m^2)$, which has no explicit dependence on the Higgs field. The inverse differential operator can be developed in a low-energy expansion in powers of $\square /m^2$ to yield a series of corrections to the SM Lagrangian: 
\be
{\cal L}_{\rm eff}=-{1\over 2} M_p^2 R+{\cal L}^{SM}+\frac{1}{2} \frac{\mu^2}{m^2}|H|^4-
g\frac{\mu M_p}{m^2} |H|^2R 
 +\frac{1}{2}\frac{\mu^2}{m^4}\left(\partial_\mu |H|^2\right)^2  + \frac{g^2}{2}\frac{M_p^2}{m^2}\;R^2 + ...
\label{Leff}
\ee
where we have neglected operators with four derivatives or more, except for the purely gravitational Ricci-squared operator. This is justified by our interest in vacuum properties or  classical inflationary dynamics,  which is usually discussed at the level of two-derivative effective actions. The $R^2$ term,  containing four derivatives, was retained on account of its potentially large coefficient,  of order $M_p^2 / m^2$. 

The first term shown in Eq.~(\ref{Leff}) as a deviation from the SM Lagrangian
is a marginal operator that shifts the SM Higgs quartic coupling. 
We find that the Higgs quartic coupling at low-energy  is
\be
\lambda = \lambda_0 - \frac{\mu^2}{2m^2}\ ,
\label{shiftedl}
\ee
in agreement with our previous definition in Eq.~(\ref{lambdaeff}).
This shift in the quartic is simply absorbed by a redefinition of the original UV coupling $\lambda_0$ and is not an observable effect. However, it is theoretically important in linking the Higgs mass (related directly to $\lambda$) to the UV behaviour of the scalar potential and can be very relevant to cure the stability problem of the Standard Model, as discussed below.

Adding the rest of operators appearing in 
 Eq.~(\ref{Leff}), we determine the following effective dynamics for the Higgs-graviton sector: 
 \be\label{effhi}
 {\cal L}_h = {1\over 2} \left(1+ \alpha\,\xi^2\,{ h^2 \over M_p^2} \right) (\partial_\mu h)^2 +{1\over 2} \mu_h^2 \,h^2 - {\lambda \over 4}\,h^4 -{1\over 2}\left( M_p^2 + \xi \,h^2 \right) \,R + {1\over 2} \,\gamma\,R^2 \;,
 \ee
where $h$ is the neutral Higgs mode. The coupling constants of irrelevant operators are calculated in terms of `microscopic' parameters  as we show below. 

The model (\ref{effhi}) is very similar to the original model of Higgs inflation introduced in \cite{shap}. In fact, it is exactly the same in the formal limit $\alpha = \gamma =0$. 
The crucial $\xi$ parameter (taken to be of order $10^4$ in the original formulation), is induced dynamically in our model and is given by
\be\label{xi}
\xi \equiv {\mu g M_p \over m^2}\; .
\ee
We can see that all it takes to generate $\xi\gg 1$ is to arrange for a mass hierarchy $\mu\simlt m \ll g M_p$ (with $\mu>0$ to get the right sign of $\xi$).

The two operators which make  (\ref{effhi}) deviate from the original Higgs inflation model are a dimension-six correction to the Higgs  field metric and a potentially large $R^2$ correction, and are  controlled  by the couplings
\be\label{constants}
\gamma \equiv {g^2 M_p^2 \over m^2} \;, \qquad \alpha \equiv {1/g^2}\;.
\ee
 We will estimate the  impact of these couplings in the extrapolated `Higgs inflation' dynamics in the next section. For now we just mention that the kinetic correction proportional to
$\alpha$ does reveal the low effective cutoff scale $\Lambda = M_p / \xi$, since  $\alpha \xi^2 /M_p^2 = \alpha/ \Lambda^2$. Furthermore, we find 
$$
\gamma \equiv {g^2 M_p^2 \over m^2 } = {m^2 \over \mu^2} \,\xi^2\;.
$$
Given that $\mu$ must stay below $m$  for low-energy stability, we see that $\gamma$ is large whenever $\xi$ is large. In fact, if we decouple the two scalars by sending $\mu / m \rightarrow 0$, the $R^2$ interaction is expected to  become more important than the non-minimal coupling (\ref{nonmin}), a situation in which  (\ref{effhi})  would support  `$R^2$ inflation'   rather than `Higgs inflation'. 

The implication is that our set up in (\ref{LJUV}) can be viewed as covering a large space of inflationary models, including standard chaotic models based on a $\phi^2$ potential, Starobinski-type inflation and Higgs inflation.  
Viewed through the lens of effective field theory, (\ref{Leff}) has a naive cutoff scale $\Lambda= M_p /\xi =m^2 /(g\mu)$. 
The phenomenological constraint $\mu < m$, together with the sub-Planckian unitarity constraint, $g\leq 1$, imply that $\Lambda$ remains  slightly above the true mass scale of new degrees of freedom, $m$,  as expected.
Hence, the transition from (\ref{effhi}) to the  two-scalar model proceeds without any strong-coupling thresholds. The UV model (\ref{LJUV})  is a partial UV completion of the effective action (\ref{Leff})  which remains weakly coupled all the way up to the Planck scale. In this way, we succeed in generating a model with the crucial ingredient of HI, namely a large value of $\xi$, out of a standard tuning of relevant parameters.

Before moving on to the analysis of the large field behaviour in our effective theory, let us make a few more comments on the  microscopic calculation of $\xi$, given in Eq.~(\ref{xi}). This result can be understood as a threshold correction for the $|H|^2 R$ operator. Above the scale $m$ only the $\phi$ field
couples nonminimally to gravity. A possible nonminimal $|H|^2 R$ coupling might be present due to radiative corrections but it will be negligible. If we look at the renormalization group (RG) evolution of $\xi$ from $M_p$ down to the EW scale, we will therefore find a negligible value from $M_p$ down to $m$, at which scale the potentially very large effect in (\ref{xi}) appears. Below $m$, the coupling $\xi$ will evolve
with its standard RG equation ($Q$ is the renormalization scale)
\be
\frac{d\xi}{d\log Q} = \frac{1}{16\pi^2} (\xi +1/6)\left[12\lambda +6h_t^2-\frac{9}{2}g_2^2-\frac{3}{2}g_1^2\right]\ ,
\ee
staying large all the way down to the EW scale.
One might worry that a large $\xi$ value could jeopardize the perturbative analysis
in the low-energy effective theory, {\it e.g.} if $\xi^2$ corrections appear in the RG for $ \xi$
at two loops. However, it is easy to see that to all orders in perturbation theory there
will be no contributions to the RG of $\xi$ higher than linear (unless hugely suppressed
by powers of $\mu_h^2/M_p^2$).

\section{Large-field Behaviour}

Armed with our partial UV completion for the Higgs inflation scenario we can now test the extrapolation procedure  implicit in standard treatments. 
The low-energy effective actions (\ref{Leff}) and (\ref{effhi}) are  nominally valid within a patch of radius $|H| \sim \Lambda$ in Higgs field space. The extrapolation procedure consists on the blind continuation towards higher values of the Higgs field, disregarding the effect of higher order corrections contained in the formal expression (\ref{forme}). On the other hand, the partial UV completion (\ref{LJUV}) defines an effective field theory free from strongly coupled operators within a patch of size $M_p$ in field space.  Outside this region, the model (\ref{LJUV}) is itself extrapolated to build an inflationary plateau. Therefore, it is the behaviour in the intermediate region, $ \Lambda \ll |H| \ll \sqrt{\xi} \,\Lambda$, that provides  a stricter test of the extrapolation procedure.

\subsection{Higgs field extrapolation}

We begin with the  extrapolation of (\ref{effhi}) into the region $h \gg \Lambda = m^2 /g\mu$. The main new dynamical feature of this domain is the large Higgs-graviton mixing induced by (\ref{nonmin}), which forces a diagonalization of kinetic terms. More precisely, the graviton fluctuations can be disentangled by passing to the Einstein frame with the metric field redefinition:  
\be
g_{\mu\nu} {\big |}_{\rm Jordan} \longrightarrow  {1\over \Omega_h} \,g_{\mu\nu} {\big |}_{\rm Einstein}  \ ,\quad \text{with}\quad 
\Omega_h \equiv 1+ \xi h^2/M_p^2 \ .
\ee
The resulting  potential in the Einstein frame is
\be
V(h)=\frac{1}{\Omega_h^2}\left[-\frac{1}{2}\mu_h^2h^2+\frac{1}{4}\lambda h^4\right]\ ,
\label{VEff}
\ee
showing the familiar flattening at large field values, with vacuum energy density $V_\infty=\lambda  M_p^4/(4\xi^2)$. This is of course exactly the same as in the original HI proposal \cite{shap}. 

As mentioned above, differences arise from the fact that, besides the crucial $\xi \,h^2 \,R$ term, decoupling $\phi$ leaves behind two other irrelevant terms not present in \cite{shap}, namely the $R^2$ term and the $(\partial_\mu |H|^2)^2$ operator appearing in Eq.~(\ref{Leff}).
Let us discuss first the impact of this last operator. As we have seen in (\ref{effhi}), it gives an $h^2(\partial_\mu h)^2$
contribution to 
the kinetic term of $h$, modifying the relation between the field $h$ and the canonically normalized field $\chi$ at large background field values. Explicitly, the kinetic
part of the effective Lagrangian in Einstein frame is
\be\label{kin}
 \frac{1}{2}\frac{(\partial_\mu h)^2}{\Omega_h^2}\left[1+(\xi+6\xi^2)\frac{h^2}{M_p^2}+\alpha\  \xi^2\Omega_h\frac{h^2}{M_p^2}\right] = \frac{1}{2}(\partial_\mu \chi)^2\ ,
\ee
where we recall that $\alpha = 1/g^2$ can be taken to be  of ${\cal O}(1)$ in order to comply with the `unitarity constraint' of the UV model. A formal limit $g^2 \rightarrow \infty$ or, equivalently $\alpha \rightarrow 0$, would give us the standard kinetic term of the original HI model studied in \cite{shap}. 

The qualitative behaviour of the field metric (\ref{kin}) determines three dynamical regimes: the low-energy one, $h \ll \Lambda$, where there is little difference between the $h$ field and the canonical field $\chi$, the intermediate regime, $\Lambda \ll h \ll \sqrt{\xi} \, \Lambda$, where we can still approximate $\Omega_h \simeq 1$, but (\ref{kin}) is already non-trivial, and finally the asymptotic or `plateau' regime, $h \gg M_p /\sqrt{\xi}$,  where $\Omega_h \simeq \xi h^2 / M_p^2$ and  the so-called Higgs-inflation takes place.

Approximating (\ref{kin}) in the intermediate domain we find
$$ 
{1\over 2} (\partial_\mu h)^2 \;(6+\alpha) \,{h^2 \over \Lambda^2} \simeq {1\over 2} (\partial_\mu \chi)^2
$$
which leads to the relation $\chi \simeq \sqrt{6+\alpha} \,h^2 /2\Lambda^2$. We conclude that the effect of $\alpha$ reduces to a mild numerical rescaling in the intermediate regime, where the model (\ref{effhi}) behaves essentially like the original version \cite{shap}, with  an   approximately quadratic potential 
\be\label{potint}
V(\chi) \simeq {\lambda\over 6+\alpha}\,\Lambda^2 \,\chi^2\,,\qquad \Lambda \ll \chi \ll M_p\;.
\ee 

The difference between $\alpha=0$ and $\alpha=1$ becomes more critical in the asymptotic `plateau' domain, which corresponds to $h\gg M_p /\sqrt{\xi}$ or, equivalently $\chi \gg M_p$. In this case we can approximate (\ref{kin}) by 
\be\label{kina}
{1\over 2} (\partial_\mu h)^2 \;\left[\alpha\,\xi + {6M_p^2 \over h^2} \right]   \simeq {1\over 2} (\partial_\mu \chi)^2
\;.
\ee
In the original HI model, $\alpha =0$, and the asymptotic field redefinition relating $h$ and $\chi$ is exponential:
\be
h^2\simeq \frac{M^2_p}{\xi}\left[\exp\left(\frac{\chi}{\sqrt{6}M_p}\right)-1\right]\ ,\quad  (\alpha=0)\ ,
\ee
so that $V(\chi)$  approached the asymptotic plateau at $\chi \gg M_p$  as an exponential
\be
V^{(\alpha=0)}(\chi) \simeq \frac{\lambda M_p^4}{4\xi^2}\left[1-\exp\left(-\frac{2\chi}{\sqrt{6}M_p}\right)\right]^2\ .
\ee
In our case, however, the extra term modifies this behaviour, leading to the simpler 
relation:
\be
h\simeq \frac{\chi}{\sqrt{\xi}}\ ,\quad  (\alpha=1)\ ,
\ee
with the potential going as
\be
V^{(\alpha=1)}(\chi) = \frac{\lambda M_p^4}{4\xi^2}\left[1-2\frac{M_p^2}{\chi^2}+...\right]\ ,
\label{VEeff}
\ee
in the asymptotic region. 
The inflationary predictions are obviously affected.  While the Hubble rate at the beginning of inflation (when the scalar field is well into the plateau region of the potential) is the same,
\be\label{hub}
H_I^2\equiv \frac{V}{3M_p^2}\simeq  \frac{\lambda M_p^2}{12\,\xi^2}\ ,
\ee
the slow-roll parameters
\be
\epsilon \equiv \frac{M_p^2}{2}\left(\frac{V'}{V}\right)^2\ ,\quad
\eta \equiv M_p^2\frac{V''}{V}\ ,
\ee
(with primes denoting $\chi$ derivatives)
will scale differently with the number of e-folds
\be
N_e=-\frac{1}{M_p^2}\int_{\chi_i}^{\chi_f}\frac{V}{V'}d\chi\ ,
\ee
with $\chi_i$ and $\chi_f$ the values of the field at the beginning and end of inflation respectively. 

For the original scenario ($\alpha=0$) the scaling of the slow-roll parameters is
\be
\epsilon\simeq \frac{3}{4N_e^2}\ ,\quad \eta\simeq -\frac{1}{N_e}\ ,
\ee
so that the scalar spectral index ($n=1-6\epsilon+2\eta$) and the tensor-to-scalar ratio ($r=16\epsilon$) are, 
\be
n\simeq 0.965\ , \quad r\simeq 0.0033\ ,
\ee
for $N_e\sim 60$, which is in good agreement with the measured values \cite{Planck}.

On the other hand, for our extrapolated model ($\alpha=1$) the scaling is instead
\be
\epsilon\simeq \frac{1}{(4N_e)^{3/2}}\ ,\quad \eta\simeq -\frac{3}{4N_e}\ ,
\label{slowscaling}
\ee
leading to
\be
n\simeq 0.973\ , \quad r\simeq 0.0043\ ,
\ee
for $N_e\sim 60$. We see that the value of the scalar spectral index is somewhat larger
than in the original scenario, moving away from the central value.

We conclude that (\ref{effhi})  shares the same qualitative properties as the standard Higgs-inflation scenario. Notably, the Einstein-frame potential has three well-separated regimes: a low-energy one dominated by the standard scale-invariant $V(\chi) \sim \lambda\, \chi^4 $ dependence; an intermediate one where the Higgs field is essentially free, $V(\chi) \sim \lambda \,\Lambda^2 \,\chi^2$; and a plateau with asymptotically constant potential. There are, of course, important differences  in the plateau region when it comes to the detailed 
predictions for the spectral index and tensor ratio, but the main questions of consistency faced by  \cite{shap} can be recreated in the effective action (\ref{effhi}), providing a good laboratory for the large-field extrapolation.

Before turning to the analysis of the UV-completed model, we must face up to one more issue. As noticed above, the $R^2$ operator comes into  (\ref{effhi}) with  a hierarchically large coefficient, $\gamma = g^2 M_p^2 / m^2 \gg 1$. This can affect the inflationary dynamics,  when the background curvature is  of order $H_I^2$. In order to estimate the relevant effects, we must first translate the $R^2$ term to the Einstein frame, resulting in a series of terms of the form
$$
\gamma \left[ R -{3\over 2}\, |\partial_\mu \,\log \Omega_h |^2 + 3 \,\square\, \log \Omega_h \right]^2\;,
$$
all of them containing two extra derivatives or one extra power of the curvature with respect to the terms previously retained in the effective Lagrangian. During inflation, the overall contribution of such terms is controlled by the Hubble scale $H_I$, so that we can expect modifications of the inflationary dynamics by a factor of order 
$$
 1+ {\cal O}\left(\gamma\,H_I^2 / M_p^2\right)\;, 
 $$ 
 which translates into relative corrections of size $g^2\, H_I^2 / m^2$ when we plug in the value of $\gamma$.  Therefore, the model (\ref{effhi}) is a good emulator of the original Higgs inflation scenario \cite{shap} provided the Hubble scale is small compared to the $\phi$-field mass, $m$. Using now  (\ref{hub}), we can translate this condition into a constraint on the Higgs coupling at the threshold:
\be\label{rco}
\gamma\, {H_I^2  \over M_p^2} \simeq { \lambda \over 12} {m^2 \over \mu^2} \ll 1\;,
\ee
a condition that can be achieved  easily for small $\lambda$,\footnote{Note, in particular that the SM running of $\lambda$ towards lower values in the UV works in the direction required by (\ref{rco}), which depends on the $\lambda$ coupling at the scale $m$.}
provided $\mu\simlt m$.
Any choice of parameters leading to a gross violation of (\ref{rco}) will correspond to a model dominated by `$R^2$-inflation' rather than `Higgs inflation'.  A simple instance leading to this situation is the $\mu \ll m$ limit, which effectively decouples the Higgs from the inflationary sector of the theory. It is interesting to notice that the type of $R^2$-inflation arising in this $\mu=0$ limit is still characterised by a `power-like' plateau similar to (\ref{VEeff}), rather than the exponential plateau characteristic of the original model \cite{starob}. In general, the generation of exponential plateaus, of either `Higgs' or `$R^2$' type, requires what amounts to the extremely unnatural $g\rightarrow \infty$  limit in our UV model.

\subsection{High energy theory}

Let us next look at the large field behaviour of the 
model (\ref{LJUV}) to see how it compares with the previous analysis.  We begin by noticing that the presence of a $\phi\,R$ coupling 
induces again a scalar-graviton mixing in the far region of $\phi$-field space. The diagonalization of this mixing is accomplished, as before, by a Weyl rescaling of the metric to the Einstein frame: 
\be
g_{\mu\nu} {\big |}_{\rm Jordan} \longrightarrow  {1\over \Omega_\phi} \,g_{\mu\nu} {\big |}_{\rm Einstein}  \ ,\quad \text{with}\quad 
\Omega_\phi \equiv 1+ 2 \phi/M_p \ ,
\ee
where we have set $g=1$ for simplicity. 
The result of this  field redefinition is the following effective action in the scalar  sector:
\be\label{effsc}
{\cal L}_{\rm scalar} =  \frac{1}{2}\sum_{i,j = h, \phi} G_{ij}\,\partial_\mu\varphi_i\partial^\mu\varphi_j - V(h, \phi)\;, 
\ee
where the Einstein-frame potential is given by 
\be
V(h, \phi) = \frac{1}{\Omega_\phi^2}\left[-\frac{1}{2}\mu_h^2 \,h^2+\frac{1}{4}\lambda_0 \,h^4+\frac{1}{2}m^2 \,\phi^2-\frac{1}{2}\mu\,\phi \,h^2\right]\ ,
\label{VEfull}
\ee
and the metric in field space is diagonal with entries \footnote{Both the potential and the field metric exhibit second order poles at the singular line $\phi_{\rm sing} = -M_p /2$. The singularity in the metric means that the singular locus lies at infinite `proper distance' in field space, whereas the positivity of the potential turns it into a  `dynamically repulsive' region.  These two facts make the singularity less problematic from the point of view of effective field theory. }
\be\label{MEfull} 
G_{\phi \phi}=\frac{1}{\Omega_\phi}\left(1+\frac{6}{\Omega_\phi}\right) ,\quad\quad
G_{hh}=\frac{1}{\Omega_\phi}\ .
\ee

A  basic observation is that (\ref{effsc}) shows no strong-coupling thresholds directly associated to the Higgs direction in field space. This is tantamount to the successful unitarization of the SM  sector up to arbitrarily high scales. The only strong-coupling threshold  visible in the model (\ref{effsc}) coupled to gravity is the usual one at the Planck scale. In particular, we can approximate $\Omega_\phi \simeq 1$ within the strip  $|\phi| \ll M_p /2$ in field space, where there  is essentially no difference between the Jordan and the Einstein frames. 

In order to compare with the extrapolated Higgs effective action of the previous section, we must project the two-field model into a single-field model. Near the SM vacuum the Higgs field is much lighter than the singlet, and the projection consists on integrating out $\phi$, an operation that obviously yields (\ref{Leff}).  Working in the static approximation, i.e. neglecting derivatives, we can substitute $\phi$ by the classical solution
\be\label{lowe}
\phi (h) \simeq {\mu \over 2m^2} \,h^2 = {h^2 \over 2\Lambda}\;,
\ee
where we have also neglected $\mu_h$ and curvature effects.   Equation (\ref{lowe}) determines a submanifold of field space which projects the two-field model into an effective theory for the Higgs field alone, consisting on Eq. (\ref{effhi}). However, as soon as we enter the region $\phi \gg \Lambda$
the role of `lightest field' is switched between the Higgs and the singlet. To illustrate this, we can compare the second derivative of the potential in the $\phi$ and $h$ directions:
$$
\partial_h^2 U = 3\lambda_0 h^2 -\mu \phi\;,\qquad \partial_\phi^2 U = m^2\;.
$$
Evaluating the ratio along the submanifold (\ref{lowe}) we find
$$
\left|{\partial_h^2 U \over \partial_\phi^2 U}\right|_{h^2 = 2m^2 \phi / \mu} = {2\phi \over \mu}\, (2\lambda_0 + \lambda) = {2\phi \over \Lambda} \,\left[1+ {\cal O} (\lambda/\lambda_0)\right] \simeq {2\phi 
\over \Lambda}\;. 
$$
In the last step we have incorporated the phenomenological constraint forcing us to work with parametrically small values of $\lambda/\lambda_0$. We see that, 
 for $\phi \gg \Lambda$, we have $\partial_h^2 U \gg \partial_\phi^2 U$ when evaluated along the line (\ref{lowe}).
Therefore, in  the region $\phi \gg \Lambda$  it is more appropriate to integrate out $h$ in favor of $\phi$. Doing this in the static approximation  induces the projection to a single-field model along the submanifold  
\be\label{trns}
h^2(\phi) =\frac{1}{\lambda_0}\left[\mu\, \phi +\mu_h^2\right]\simeq  \frac{1}{\lambda_0}\mu \,\phi = 2\Lambda \,\phi\,\left(1- {\lambda\over \lambda_0}\right) \ ,
\ee 
(we can again neglect $\mu_h^2$ in this region of field space).  Notice that the submanifold (\ref{trns}) differs slightly from (\ref{lowe}), by the appearance of the ${\cal O}(1)$  factor $(1-\lambda/\lambda_0)$.  

Substituting (\ref{trns}) into the kinetic term of (\ref{effsc}) gives us the effective metric on the submanifold (\ref{trns}):
\be\label{efme}
{1\over 2} G_{\phi\phi} \,(\partial \phi)^2 + {1\over 2} G_{hh}\, \left(\partial \sqrt{\mu\phi/\lambda_0}\right)^2 = {1\over 2}\left(G_{\phi\phi} + G_{hh} {\mu \over 4\lambda_0 \phi} \right) \;(\partial \phi)^2\;.
\ee
Since $G_{hh} \simeq 1$ and $ G_{\phi\phi} \simeq 7$ as long as $\phi\ll M_p$, we find that the field-space metric induced on the submanifold (\ref{trns}) is essentially trivialised by the field $\phi$ in the intermediate domain 
 $\Lambda\ll \phi \ll M_p$, i.e. the canonical field is given  simply by $\chi \simeq \sqrt{7}\,\phi$. This region maps to the `intermediate region' in the extrapolated Higgs model of the previous section, and the resulting effective potential for the canonical field is given approximately by
\be\label{intep}
V(\chi) \simeq { \lambda \,\Lambda^2 \over 7} \left(1- {\lambda \over \lambda_0}\right)\, \chi^2\;,\qquad \Lambda \ll \chi \ll M_p\;,
\ee
which coincides with the potential (\ref{potint}) of the extrapolated model (evaluated for $\alpha=1$), up to the ${\cal O}(1)$ correction factor $(1-\lambda/\lambda_0)$. Notably, the light mode behaves as an essentially free field with mass of order $\sqrt{\lambda} \Lambda$ in both cases.

As we continue to the trans-Planckian region $\phi \gg M_p$, the canonical light field continues to be well approximated by the trajectory (\ref{trns}), but the field metric becomes non-trivial in the asymptotic region. Evaluating (\ref{efme})  in the asymptotic region  $\phi \gg M_p$, we find
$$
{1\over 2} (\partial \chi)^2 \simeq {1\over 2} G_{\phi\phi} (\partial \phi)^2 \;,
$$
leading to a canonical field
$$
\chi \simeq \sqrt{2M_p \phi}
$$
and an associated asymptotic potential 
$$
V(\chi) \simeq V_\infty \left[1- 2 {M_p^2 \over \chi^2} + {\cal O}(M_p^4 / \chi^4) \right]
\;.
$$
The asymptotic vacuum energy is given by 
\be
V_\infty=\frac{\lambda M_p^2 m^2}{8\lambda_0}=\frac{\lambda M_p^4}{4\xi^2}\left(1-\frac{\lambda}{\lambda_0}\right)\ ,
\ee
which, compared to the value from the effecive theory (\ref{VEeff}), indeed shows the same overall correction factor as (\ref{intep}).

We conclude that the rough features of inflationary dynamics, as well as  the classical properties of the lightest fields in the intermediate region, are qualitatively well described by  the extrapolated `Higgs inflation'. This is happening despite the fact that the dynamics is actually dominated by $\phi$ beyond the $\Lambda$ threshold. The main difference between the models, as described in the static approximation for heavy fields, is the emergence of a correction factor $1-\lambda/ \lambda_0$ in the energy density, which finds its origin in the slight mismatch between the submanifolds (\ref{lowe}) and (\ref{trns}).

\subsubsection{Detailed inflationary dynamics}

 The previous `broad brush' analysis of the inflationary dynamics can be further substantiated by a more careful treatment of the motion in field space. The slow-rolling fields move along trajectories that satisfy the differential equation
\be
\frac{dh}{d\phi}= \frac{G_{\phi\phi}}{G_{hh}}\left(\frac{\partial V/\partial h}{\partial V/\partial\phi}\right)\ .
\ee
For illustration, the integral curves of this equation are shown in Figure~1.
In the large field region, the trajectory that asymptotically reaches the valley of the potential can be obtained analytically in inverse powers of $\phi$ as
\be
h^2(\phi)= \frac{\mu \phi}{\lambda_0} +
\frac{\lambda M_p m^2}{4\lambda_0^2\phi}
\left[1-\frac{14\lambda_0 M_p +\mu}{4\lambda_0\phi}+{\cal O}(M_p^2/\phi^2)\right]\ .
\label{valleyline}
\ee
This particular trajectory, the inflationary attractor, is also shown in Figure~1 as an red line. Deep along the plateau, it is well approximated by the submanifold (\ref{trns}), which was derived by freezing the classical dynamics of $h$.

\begin{figure}[t]
$$\includegraphics[width=\textwidth]{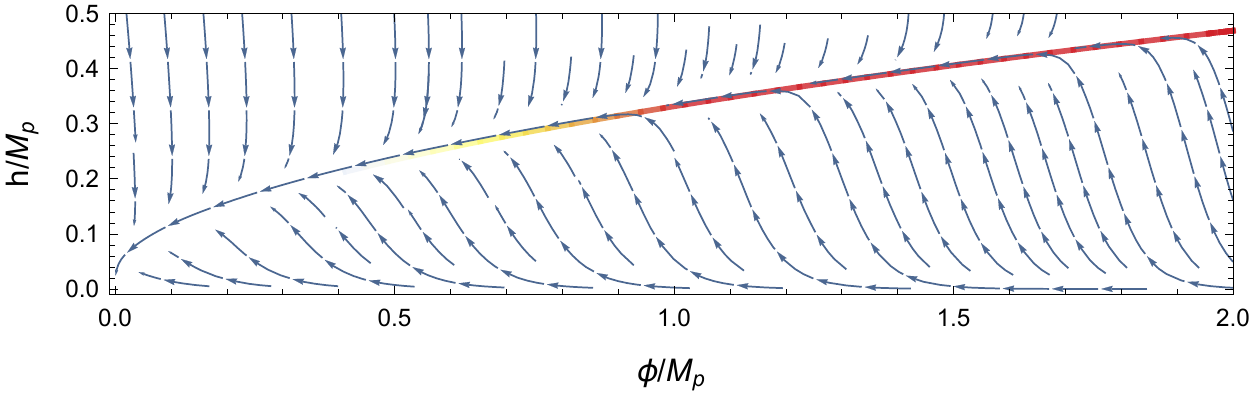} $$   
\begin{center}
\caption{\emph{Illustaration of the lines of flow of a slow-rolling field in the potential $V(\phi,h)$ of Eq.~(\ref{VEfull}). The red line shows the inflationary trajectory
(\ref{valleyline}) along the potential valley (with the transition to yellow indicating that slow-roll no longer holds).}\label{fig:roll}}
 \end{center}
\end{figure}

We can calculate the slow-roll parameters in this two-field scenario
using the standard generalization to the multi-field case:
\be
\epsilon \equiv \frac{M_p^2}{2V^2}G^{ij}\frac{\partial V}{\partial \varphi_i}\frac{\partial V}{\partial \varphi_j}\ ,\quad
\eta\equiv \text{Min Eigenvalue} \left[\frac{M_p^2}{V}G^{ik} \frac{\partial^2 V}{\partial \varphi_k\partial \varphi_j}\right]\ ,
\ee
where $G^{ij}$ is the matrix inverse of $G_{ij}$. Along the valley trajectory 
(\ref{valleyline}) we get
\be
\epsilon= \frac{M_p^3}{\phi^3}+{\cal O}(M_p^4/\phi^4)\ ,
\quad
\eta= -\frac{3M_p^2}{\phi^2}+{\cal O}(M_p^3/\phi^3)\ ,
\ee
while the number of e-folds is given by the integral
\be
N_e=-\int_{\phi_i}^{\phi_f}\frac{1}{2\epsilon V}\frac{dV}{d\phi}d\phi\ ,
\ee
where $V$ is understood  here as the potential along the valley, that is,
$V(\phi)=V[h(\phi),\phi]$, with $h(\phi)$ as given by Eq.~(\ref{valleyline}).
One can easily obtain that the scaling of $\epsilon$ and $\eta$ with the number of e-folds is exactly the same found in the low-energy effective theory, given in Eq.~(\ref{slowscaling}).

\subsection{Potential Stability}

We close this section by addressing the issue of potential stability, that
poses a serious threat to the scenario of Higgs inflation. As is well known, for the central experimental values of the Higgs and top masses, $m_h$ and $m_t$, (and assuming no BSM physics up to the Planck scale), the SM Higgs potential develops an instability at large field values \cite{mhstab} making the electroweak vacuum metastable. Although such metastability is mild and the lifetime of the vacuum is extremely large compared with the age of the Universe, the viability of Higgs inflation does require a positive potential at large field 
values.\footnote{For another recent way to overcome this problem and to access the same parameter region, see \cite{shaplatest}.} However, stability of the potential at large scales  requires experimentally disfavored values of $m_t$ and/or $m_h$ (even allowing for a larger error in the determination of $m_t$ \cite{djouadi})
disfavoring also the possibility of Higgs inflation \cite{Salvio}.

A simple, generic and very efficient cure of the potential instability was proposed in Ref.~\cite{threshold}. The idea is to add to the SM a singlet field $S$ with a large vacuum expectation value and coupled to the Higgs as $\lambda_{HS} |H|^2 S^2$. The effective theory below the singlet mass is SM-like with a negative threshold effect on the Higgs quartic coupling that makes the Higgs mass lower than it would be without the singlet coupling. In the low-energy effective theory the
Higgs mass looks dangerously light for stability, but the UV-complete theory above the singlet threshold does not suffer any stability problem. This simple mechanism can be implemented
in many different models with singlets (see \cite{threshold} for a few relevant examples) and it was indeed applied in  \cite{threshold}
to the unitarized-Higgs-inflation scenario of \cite{gianlee}.

In the model we have presented in this paper we have a similar kind of threshold correction to the Higgs quartic coupling and with the right sign to help stabilizing the potential at large field values,
see Eq.~(\ref{shiftedl}), even though the details of the stabilization mechanism are not those of \cite{threshold} (the threshold effect does not depend on the singlet vacuum expectation value and
there is not a $\lambda_{HS}$ coupling). The stability conditions for
the potential of Eq.~(\ref{VEfull}) are simply
\be
m^2> 0\ ,\quad \quad \lambda >0 \ ,
\ee
which come from requiring stability along the directions $\phi$ and $h^2\simeq \mu\phi/\lambda_0$, respectively. 
The last condition is in fact the same we would impose on the low-energy effective theory so that there is no tree-level gain from the threshold effect concerning the stability conditions. There is however
a significant gain at the loop level as the new degree of freedom $\phi$ changes the renormalization group evolution of the Higgs quartic coupling above the $\phi$ threshold scale. We can see this by writing
\be
\frac{d\lambda}{d\log Q} = \beta_\lambda^{SM}  +  \frac{1}{2\pi^2}(\lambda_0-\lambda)(\lambda_0+2\lambda)\ ,
\ee
that shows how the running of the quartic coupling with the renormalization scale $Q$ receives sizeable positive contributions from the singlet. This effect is enough to stabilize the potential and allow to extend the scenario of Higgs inflation to the experimentally preferred range of
Higgs and top masses provided the mass of the field $\phi$ is below the 
SM instability scale ($\Lambda_i\sim 10^{11}$ GeV for the central values of $m_h$ and $m_t$). For a given $\xi$, this condition translates into an upper bound
on $\mu$,
\be
\mu < \xi \frac{\Lambda_i^2}{g M_p}\ ,
\ee
that is easy to satisfy.

\section{Discussion}

The model presented in this paper realizes a simple (partial) UV completion of a Higgs-inflation (HI) scenario. Besides the Higgs field, $h$, the model contains an additional scalar field, $\phi$, coupled linearly to the Ricci curvature scalar with an strength of ${\cal O}(1)$  in $M_{ p}$ units. 
Apart from the Einstein-Hilbert term, there are no other irrelevant operators, so unitarity is preserved below the Planck scale. In the complete theory, inflation is driven essentially by the $\phi$-field. The effective theory, obtained by integrating-out $\phi$, is very similar to the `standard' HI model postulated in ref.\cite{shap}. Remarkably, the abnormally large dimensionless parameter $\xi$ of that HI model is simply generated here as a prosaic ratio of mass scales. 

Beyond particular characteristics, our model illustrates the fact that HI, understood as a scenario in which inflation is solely driven by the dynamics of the Higgs field, can be simply a mirage effect from the true inflationary process, in which the Higgs field might play a rather secondary role. This is illustrated  in Figure 2, a qualitative rendering of $(h,\phi)$ field space showing the comparative accuracy of the extrapolation from the low-energy theory.  
The submanifold of configuration space determined by Eq.~(\ref{lowe}),  which is selected by the extrapolation procedure, is represented by the dashed line.
 On the other hand, the continuous parabolic line represents the  approximate `valley' of Eq.~(\ref{trns}), determined by integrating out the Higgs  mode in the static approximation. This approximation is very good when it comes to slow-roll dynamics, since (\ref{trns})  reproduces the first term in the large-field expansion (\ref{valleyline}) of the exact inflationary attractor. 
The slight difference between the two submanifolds quantifies the `fidelity' of the mirage provided by the extrapolation of (\ref{effhi}) and is responsible for the offset factors of order $1-\lambda/\lambda_0$, found in the computation of the effective potential.

The global perspective offered by  Figure 2 implies that inflation is mostly given by $\phi$-field dynamics, and yet  we have seen in Section 3 that the HI picture gets the qualitative features of the potential essentially right, particularly for small values of $\lambda / \lambda_0$. 
All these facts are likely to generalise beyond the particular model presented here. Actually, one can expect that any inflation model in which the Higgs field evolves during the inflationary process (due to some interaction with the inflaton, even if small), will show up in the effective theory as a HI scenario, where the higher order operators play a prominent role.

\begin{figure}[t]
$$\includegraphics[width=14cm]{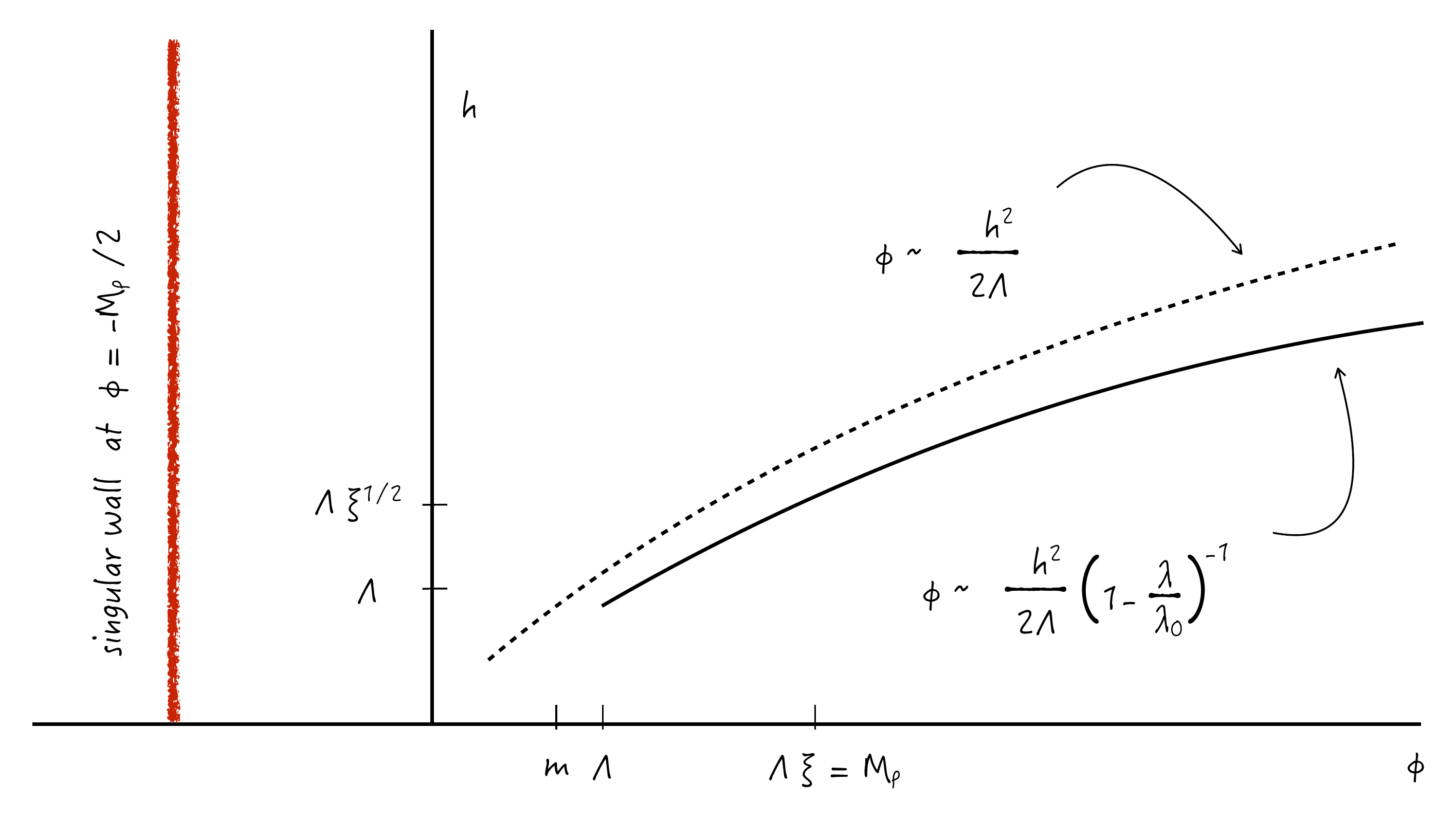} $$   
\begin{center}
\caption{\emph{Global view of the $(h, \phi)$ field space. The single-field projection  implied by the extrapolated Higgs theory is represented by the dashed line. It differs from the approximate two-field valley  (full line) for $\phi \gg \Lambda$. In the plateau region, $\phi > M_p$, the full line is a good approximation to the slow-roll attractor trajectory.  }\label{fig:media}}
 \end{center}
\end{figure}

The restriction to two-derivative effective actions is an implicit feature of the extrapolation procedure. This restriction is actually equivalent to the projection of the dynamics onto a one-dimensional submanifold of field space, such as the ones featuring in Figure 2, obtained by applying the static approximation to either $h$ or $\phi$ fields.
% From this point of view, the mirage arises when a given submanifold is used beyond the region where it gives a good approximation to the physics of interest. 

On the other hand, the non-local action (\ref{forme}) for the Higgs field, containing an infinite tower of higher derivative operators, should give an exact account of the classical dynamics of the full theory (\ref{LJUV}). This suggests that any mirage could be improved by going beyond the two-derivative level. 
As an example of this general phenomenon, let us consider an extremely simple situation arising in the $g\rightarrow 0$ limit of our model, i.e. the same as  Eq.~(\ref{LJUV}), but with the non-minimal coupling to gravity turned off. The scalar sector is now given by 
\be
\label{Lsimple}
{\cal L}_{\rm scalar}= \frac{1}{2}(\partial_\mu \phi)^2 + \frac{1}{2}(\partial_\mu h)^2 - \left[\frac{1}{2}m^2\,\phi^2 -\frac{1}{2}\mu\, \phi\, h^2\ + \frac{1}{4}\lambda_0\, h^4\right]\ ,
\ee
where $h$ is the (real) neutral Higgs component. We have obviated the usual Einstein term, $-M_p^2 \,R/2$, and the SM mass term, $- \mu_h^2 h^2 /2$, that play no role in the present discussion. 
Once again, we assume  $m\gg v$ so that, for small values of $h$ ($\sim$ EW scale), the $\phi$ field decouples and we recover the Standard Model, with the effective quartic coupling given by Eq.~(\ref{lambdaeff}), that is,  $\lambda \equiv \lambda_0 - \mu^2/(2m^2)$.
Similarly, for large field values the potential has a valley in the $h^2(\phi) \simeq \mu \phi/\lambda_0$ direction. However, along this valley the potential does not develop a plateau but rather increases  quadratically, essentially along the $\phi$-field direction, with an effective mass, 
\be
\label{mtilde}
{\widetilde m}^2 \simeq m^2-\frac{\mu^2}{2\lambda_0}= 2\frac{\lambda m^4}{\mu^2}\left(1-\frac{\lambda}{\lambda_0}\right)\ .
\ee
For convenience we assume (as we have done throughout the paper) ${\widetilde m}^2\ll m^2$, which is equivalent to $\lambda\ll \lambda_0$, implying $\lambda_0 \simeq \mu^2/2m^2$.
The direction orthogonal to the valley is mostly $h$ field, with a large effective mass of order $2 \lambda_0 h^2$. So, in the large field region  the model supports chaotic inflation, mainly along $\phi$. 

Let us now examine the single-field description obtained by the (classical) integration-out of the $\phi$ field. Plugging the equation of motion for $\phi$ in (\ref{Lsimple}), we get the $g\rightarrow 0$ limit of (\ref{forme}):
\be
\label{Lsimpleeff}
{\cal L_{\rm nl}}= \frac{1}{2}(\partial_\mu h)^2 + \frac{1}{8} h^2 \frac{\mu^2}{m^2+\square}h^2 - \frac{1}{4}\lambda_0\, h^4\ .
\ee
The solutions of the equation of motion derived from this non-local Lagrangian are simply those of the complete Lagrangian (\ref{Lsimple}) projected into the $h$ axis of field space. 
In particular, ${\cal L_{\rm nl}}$ describes the dynamics of the lightest state, that can be extracted by expanding in powers of $\square/m^2$, 
\be
{\cal L}_{\rm eff}= \frac{1}{2}(\partial h)^2\left(1+\frac{\mu^2}{m^4}h^2\right)-\frac{1}{4}\lambda\, h^4 \ +\ \cdots
\label{Lsimpleeff2}
\ee
where the dots denote higher-order terms in ${\square}/{m^2}$. Note that this two-derivative  effective Lagrangian is that of Eq.~(\ref{Leff}) after switching off the non-minimal gravitational couplings. As a matter of fact, the Lagrangian (\ref{Lsimpleeff2}) represents a remarkably simple (and succesful) scenario of Higgs inflation. In the small-field regime, $h$ is canonically normalized and 
the Lagrangian  describes  just the ordinary SM, as expected. However,  when extrapolated to the large-field regime, $h\gg m^2/\mu$, the canonically normalized field reads
\be\label{lfa}
\chi\simeq \frac{\mu}{2m^2}h^2\ , 
\ee
so that
\be
{\cal L}_{\rm eff}= \frac{1}{2}(\partial \chi)^2-\frac{1}{2}m_\chi^2\ \chi^2 \ +\ \cdots
\label{Lsimpleeff3}
\ee
with 
\be
\label{mchi}
m_\chi^2=\frac{2\lambda m^4}{\mu^2}\ .
\ee
This  has the same form as the chaotic inflationary potential derived from the complete Lagrangian (\ref{Lsimple}), except that 
$m_\chi^2$ replaces $\widetilde m^2$,  the two differing by the, by now standard, correction factor $(1-\lambda/\lambda_0)$, which measures the accuracy of the mirage extrapolation.

The simplicity of this model allows us to track the leading ${\cal O}(\lambda/\lambda_0)$ corrections that explain the difference between $m_\chi$ and ${\widetilde m}$. 
Conceptually, the non-local  Lagrangian (\ref{Lsimpleeff}) contains the same information as the `UV-model' with two scalar fields (\ref{Lsimple}), at least when considering classical field dynamics.  In order to make this more explicit,  let us pick  a classical solution $\chi_c$ of the large-field effective Lagrangian (\ref{Lsimpleeff3}), i.e. one that satisfies 
$$
\square\, \chi_c = - m_\chi^2 \,\chi_c\;. 
$$
The leading four-derivative correction  coming from (\ref{Lsimpleeff}) is 
$$
{\mu^2 \over 8m^6} \,h^2 \,\square^2 \, h^2 \simeq {1\over 2m^2} \,\chi\, \square^2 \,\chi\;,
$$
where we have used the large-field approximation (\ref{lfa})  to extract the operator in terms of the canonical field $\chi$. Evaluating
this term as a perturbation to the on-shell  value of the  Lagrangian (\ref{Lsimpleeff3}) we find
\be\label{neon}
{\cal L'}_{\rm eff} [\chi_c] = {1\over 2} (\partial \chi_c)^2 - {1\over 2} m_\chi^2 \,\chi_c^2 + {1\over 2m^2} \chi_c \,\square^2 \,\chi_c = {1\over 2} (\partial \chi_c)^2 - {1\over 2} {\overline m}_\chi^{\,2} \,\chi_c^2\;,
\ee
where 
$$
{\overline m}_{\chi}^{\,2} = m_\chi^2 \left(1- {m_\chi^2 \over m^2}\right) = m_{\chi}^2 \left[1- {\lambda \over \lambda_0} + {\cal O}(\lambda^2)\right]\;.
$$
Hence, ${\overline m}$ and ${\widetilde m}$ are found to coincide up to ${\cal O}(\lambda^2)$ effects. In this way we see that one may recover the ubiquitous $1-\lambda/\lambda_0$ factor when we keep track of higher-derivative corrections.

Of course, in passing from (\ref{Lsimpleeff2}) to (\ref{Lsimpleeff3}) we have extrapolated the Higgs field beyond the cut-off $\Lambda\sim m^2/\mu$, that can be read from the non-renormalizable operator in (\ref{Lsimpleeff2}). This is not different from the practice in conventional HI models. Clearly, the inclusion of generic additional higher-order operators would render the theory out of control for such large field values. From the pure low-energy perspective the absence of those operators would look like a mysterious conspiracy. 
However, the effective origin of the low-energy theory allows to understand their absence: namely, from Eq. (\ref{Lsimpleeff}) one sees that all the additional higher-order operators have the structure $(\mu^2/m^2)h^2 (\square/m^2)^n h^2$, which in the  small $\square/m^2$ regime give small (but not necessarily negligible) corrections to the effective Lagrangian (\ref{Lsimpleeff2}).

The simple exercise just discussed also illustrates the limitations of playing just with the effective theory. We have noted that the inclusion of the additional higher-order operators derived from the (classically) exact expression (\ref{Lsimpleeff}) imply an ${\cal O}(\lambda/\lambda_0)$ shift of the effective mass in the inflationary regime. From the low-energy point of view it is not possible to guess the size of these additional operators, which, depending on the size of $\lambda/\lambda_0$, might be necessary to extract accurate quantitative predictions. This happened also for the model discussed in depth along this paper, and it is likely to be a generic property of HI models, understood as effective theories. The details of the UV completion seem necessary to extract robust quantitative predictions. 

The `mirage' interpretation of Higgs inflation which we have presented here is likely to suffer additional `blurring' when the impact of radiative corrections is taken into account, a problem of immediate interest in the light of our analysis. Furthermore, it should be stressed that our `UV models' shed no light on the actual mechanism generating the inflationary plateau. The reason is of course the trans-Planckian nature of inflation in all the models under consideration in this paper.

%%%%%%%%%%%%%%%%%%%%%%%%%%
\section{Acknowledgments}
\label{ackn}
J.E.-M. and J.R.E. thank IFT-UAM/CSIC for hospitality and partial financial support.
This work has been partly supported by the Spanish Ministry MEC under grants FPA2013-44773-P, FPA2012-32828, and FPA2011-25948, by the Generalitat grant 2014-SGR-1450 and by
the Severo Ochoa excellence program of MINECO under the grants SO-2012-0234 and SO-2012-0249.
The work of J.E.M. has been supported by the Spanish Ministry MECD through the FPU grant AP2010-3193.

%%%%%%%%%%%%%%%%%%%%%%%%%%%%%%%%%%%%%%%%%%%%%%%%%%%%%%%%%%%%%%%%%%%%%%%%%%%%%%%%%%%%%%%%%%%%%%%%%%%%%%%%%%%%%%%%%%%%%%%%%%%%%%%%%%%

\end{document}